# Risk management in design process for Factory of the future

Guangying JIN, Séverine SPERANDIO and Philippe GIRARD

IMS-LAPS, UMR 5218, University of Bordeaux, France,

(email:guangying.jin@u-bordeaux.fr)

*Abstract—* the current globalization is faced to the rapid development of product design process with the different structure of the actor relationships in the process. Currently, the risk in the failure relationship among different actors in the project is shaped by the complexity towards the future all kinds of challenges. When it comes to the interdependent failure effect, the risk management for future organization structure in design process will be much more complex to grasp. In order to cope with adaption of Product-Process-Organization (P-P-O) model for industry of the future, we propose a risk management methodology to cope with this interdependent relationship structure. The main objective of this research is to manage the risks, so that the project manager can find the priority order of all the actors' total effect to the project with the consideration of interdependent failure affection, and according to the order, project manager can release corresponding respond measures.

## I. INTRODUCTION

The design environment is defined as the context in which the project manager wants to place the designers in order to achieve the design objectives [1]. Relatively to the product, design environment includes nature, complexity and status in process of the product [1].

Traditionally, the performance of the industrial design always concerning about the product indicators and the result of the design activities depends on the product models and design process model [2]. However, we ignored the importance of the connections among the product, process and organization [3], [4], [5]. The organization can be seen as the social entities that are goal-directed, are designed as deliberately structured and coordinated activity systems, and are linked to the external environment [6], [7], [8].

For the Product-Process-Organization (P-P-O) model [9], it is a model to describe the design system, which not only integrate elements linked with the product, process and organization but take into account clearly the human aspects [10].

Traditionally, the relationships among actors in the industrial design organization always the hierarchical structure (Fig. 1). However, in a global and Internet-driven economy, the rapid movement of people and goods across borders leads to the necessity to have a design process very reactive and adjustable. In this case, traditional hierarchical organizational structure tends to adapt slowly to changing needs, decision making and increase the communication barriers.

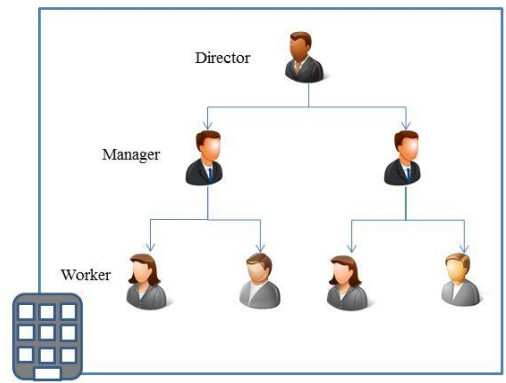

Fig. 1. Hierarchical organization structure

In the new concept of the industry 4.0 [11], [12], the hierarchical organization structure should be changed to the horizontal integration through value networks (Fig. 2).

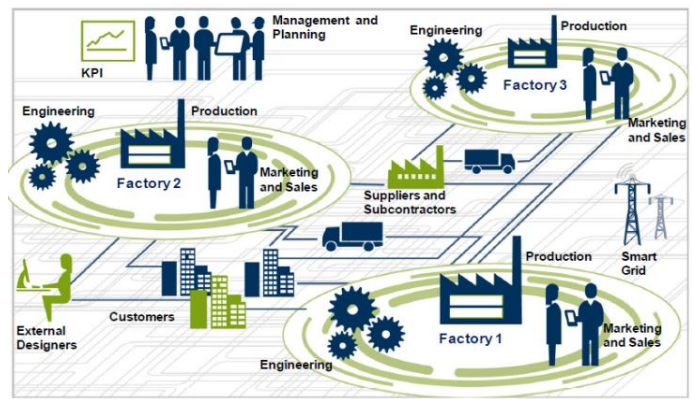

Fig. 2. Horizontal value network [13].

In here, the linking does not stop at the internal boundaries of the company but also includes suppliers, customers and other external partners (External Designers in the Fig. 2), transforming the value chain into a value network.

Hence, according to the innovative ideas in industry 4.0, third industrial revolution [14], factory of the future [15], [16], Peer-to-peer network [17] and collaborative network [18], the future organization will be the horizontal and point-to-point structure (Fig. 3).

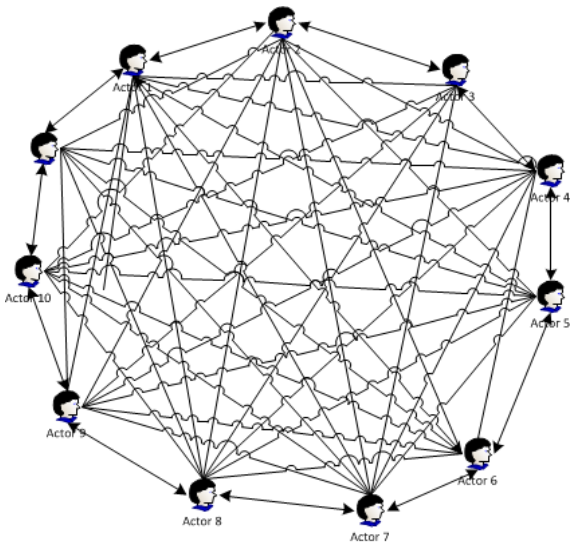

Fig. 3. Future organization structure.

In the future organization structure, all the actors can connect with any parts of them without a central intermediary. In other words, every actor will be interdependent with each other. The actors will be collaborated more and more closely, and the structure can adapt global competitive pressure and product development process complexity due to the fact that there are much more flexible and fast communication and information interaction environment.

At the organizational level, various factors may contribute to an increase in incidents and accidents, and a study by the Institute of Nuclear Power Operations showed that at least 92% of the underlying causes of accidents were caused by people [19]. Meanwhile, present technological development towards high hazard systems requires a very careful consideration by designers of the effects of "human errors" which are commonplace in normal, daily activities, but unacceptable in large-scale systems [20].

Therefore, in design process, it is very important to manage the failure risk, which is caused by the designers. Especially in the interdependent point-to-point structure for the future organization, the personal failure, will affect other actors depends on the intricate relationships with others. From here, the risk of failure will be the personal failure with the combination of a series related direct effect or inverse effect failure, which is the phenomenon of domino effect. In the future organization, the independent risk seldom exists in reality. These kinds of interdependent risks will increase the difficulty of project risk management and decrease the utility of project risk approaching methods. Therefore, analysing the affection of hidden risk for the future organization structure is urgent for the future industrial design.

In here, the main objective of this research is to manage the failure affection level for the actors, who promote the failure, in future organization structure with the point-to-point interdependent relationship consideration. From here, we can take into account the corresponding failure affection instead of isolated risk analysis for the failure promote actor itself. The result of the failure affection level is to calculate the total failure risk level order of all the actors, and let project manager to respond different actions depends on the risk level. Therefore, we will propose a risk management methodology to let the P-P-O model adapts the factory of the future.

The remainder of this paper is organized as follows. Section2 recalls the state of the art of the related work about risk management methodologies. In Section3, we introduce the risk management method for future organization structure. A summary and an outlook conclude this paper will be illustrated in Section 4.

II. RELATED WORK FOR RISK MANAGEMENT FOR FUTURE ORGANIZATION STRUCTURE

In the over 20 years, there are many types of risk management methodologies for the different kinds of organizations. Interactions-based Risk Clustering Methodologies and Algorithms is an additional clustering methodology, which could take into account the interactions between risks, in terms of existence and strength [21]. It will, firstly, identify the possible risk interactions with the binary matrix representation, and then, transformed the matrix to a numerical one using the principles of AHP (Analytic Hierarchy Process) [22]. In the part of risk interactions, however, it is just separately considering about the direct effect, while there also have indirect effect (one risk will affect anther actor through the third actor) between risks. An optimization model [23] is considered the risk interdependence and its two directions for selecting risk response strategies. In here, even though, he provides an approach to measuring risk interdependence with the consideration of strength of risk interdependence but it is also impossible for this method to confront with the problem of indirect effect between risks. A risk management methodology [24] for project risk dependencies is proposed in 2011. In this methodology, also describe the different possible of the risk dependencies between two risks. However, for the analysis of the risk, this method only considers about the probability and impact of the risk while the detection of the risk is a very important attribute for the risk too. In the method of Using Interconnected Risk Maps [25] and A System of System Approach [26] also have the same problem about the consideration of detection of the risk.

Hence, depends on the defects above, in this paper, the risk management methodology for future organization structure will consider about the combination of direct effect and indirect effect between actors' risks, severity of the risk, occurrence of the risk, and detection of the risk.

### III. RISK MANAGEMENT FOR FUTURE ORGANIZATION STRUCTURE

To describe the whole process of risk management methodology for future organization structure, we can introduce the risk management process in ISO 31000:2009 [27], [28], [29]. The ISO 31000 is a standard aims to provide organizations with guidance and a common platform for managing different types of risks, from many sources irrespective of the organizations size, type, complexity, structure, activities or location [30]. In here, the main approach for the risk management process model is to provide a generic risk management framework that is applicable to different industries and different problem scopes.

Apart from that, to evaluate and assess the risk of the future organization model, we should also introduce the risk management techniques Failure Mode and Effect Analysis (FMEA) [31], which have been developed over more than 40 years. The FMEA analysis, which is a modern and numerical risk analysis method, is a quality tool that is used to determine the potential failures of a product or system and to identify their reasons and effects [32]. Preventing the risks in process and product problems before they occur is the purpose of FMEA [31].

Depends on the Risk management process in ISO 31000 and FMEA analysis methods, we can describe the whole risk management process for future organization structure as Fig. 4.

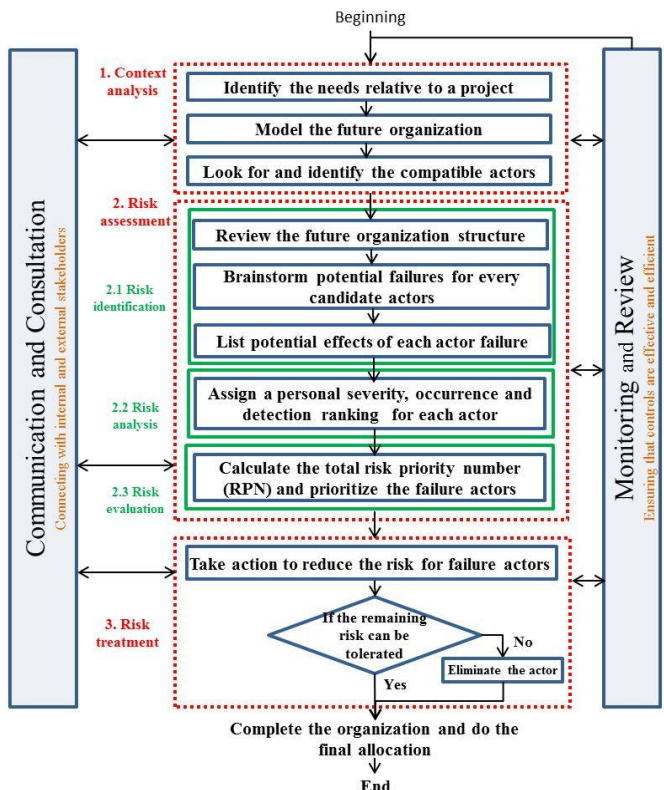

Fig. 4. Risk Management Process for future organization structure

Firstly, project manager should define the context, which depends on the communication and consultation with external and internal stakeholders. In here, firstly, project manager should identify the needs relative to the project. After that, depends on the needs and requirement of the project, we can make the future organization model which is the horizontal point-to-point structure (Fig. 3). Because of the point-to-point structure, the environment of communication and collaboration among actors will be much more closely and frequently. Meanwhile, all the actors in the model belong to the assigned project. Therefore, to exactly describe the model, the project manager should collect all the internal and external information, which from project constraints and stakeholders, related to actors' ability and relationship among actors. In addition, project manager can look for and identify the compatible actors depends on the future organization structure and needs of the project.

After that, in the step of risk assessment, project manager should review the future organization model. In the future organization, the independent risk seldom exists in reality. Personal failure in one actor will affect other corresponding actors, which can be the direct and indirect effect with the power relationship [33], and also the effect can be favourable and unfavourable. The unfavourable effect will increase the expected loss by increasing the impact of the other risk, while the favourable effect will reduce the expected loss by impact of the other risk [23]. The failure risk relationship between two actors can be seen as the Fig. 5.

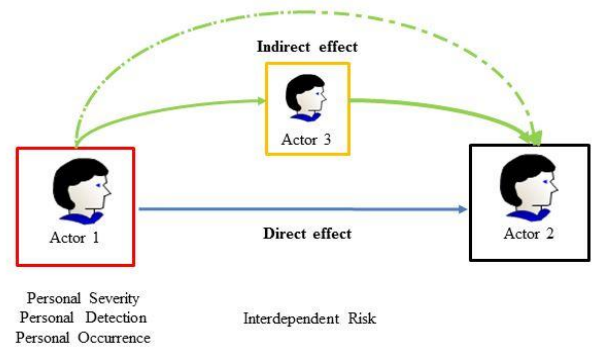

Fig. 5. Failure risk relationship between actors

In the Fig. 5, we can find that the failure effect relationship between two actors can be divided as direct effect (direct arrow Actor1 to Actor2) and indirect effect (dotted arrow from Actor1 to Actor2). In here, the direct effect means the existing failure in the actor 1 will cause the Actor2 produce failure. In addition, the indirect effect means that the failure in the Actor 1 will also influence Actor2 via transitional actors (Actor3), which just like the domino effect. Therefore, when we considering about the influence of the risk in one actor, we should not only consider about the personal failure but also think about the direct and indirect influence failure to other actors. The factors to determine the risk description of a failure and its effects are Severity, Detection and Occurrence. Also, the risk type will be divided as two parts Personal risk (Personal factors in Actor1) and Interdependent risk (Interdependent risk for direct and indirect dotted arrow). The total risk level for the failure existing actor (Actor 1) can be the total value of personal risk with the interdependent risk. The evaluation criteria rank for the severity, detection and occurrence can be seen as the Table 1 to Table 3.

TABLE I. SEVERITY EVALUATION CRITERIA

| Effect | Description | Rank |
|---|---|---|
| Negative High effect | Failure to meet the Functional definition in design process | 3 |
| Negative Medium effect | Failure to meet the Organic definition in design process | 2 |
| Negative Low effect | Failure to meet the Operational definition in design process | 1 |
| No effect | No effect to the design process | 0 |
| Positive Low effect | Success to meet the Operational definition in design process | -1 |
| Positive Medium effect | Success to meet the Organic definition in design process | -2 |
| Positive High effect | Success to meet the Functional definition in design process | -3 |

TABLE II. DETECTION EVALUATION CRITERIA

| Opportunity for Detection | Description | Rank |
|---|---|---|
| No detection opportunity | No current detection method; Cannot detect or is not analysed | 5 |
| Not likely to detect at any stage | Company has a weak detection capability | 4 |
| Moderate to detect at any stage | Company has a moderate detection capability | 3 |
| Easy and comprehensive to detect at any stage | Company has a strong detection capability | 2 |
| Detection not applicable; Failure prevention | Failure cause or failure mode cannot occur because it is fully prevented. | 1 |

TABLE III. OCCURRENCE EVALUATION CRITERIA

| Likelihood of Failure | Description | Rank |
|---|---|---|
| Very High | Failure is inevitable with new design, new application | 5 |
| High | Frequent failures associated with similar designs or in design simulation and testing | 4 |
| Moderate | Occasional failures associated with similar designs or in design simulation and testing | 3 |
| Low | No observed failures associated with almost identical design or in design simulation and testing | 2 |
| Very Low | Failure is eliminated through preventive control | 1 |

From the Table I, we can find that the effect of the failure can be positive and negative to the design process. In here, the positive value of failure means that the behaviours of the actors will have a strong inclined to make the failure (error, mistakes) and cause a negative effect to the design process while the negative value of failure means the behaviours of the actors will have a strong inclined to defeat this failure and make a positive effect to the design process. Apart from that, the divide of the occurrence of the failure rank in the Table III depends on the experience of previous associated design projects.

After review the future organization structure, project manager should brainstorm potential positive and negative failure for every actor in the project, and with deduce of the effect for these failures. The common potential positive and negative failure for the designer and corresponding effect for these failures can be seen as the Fig. 6.

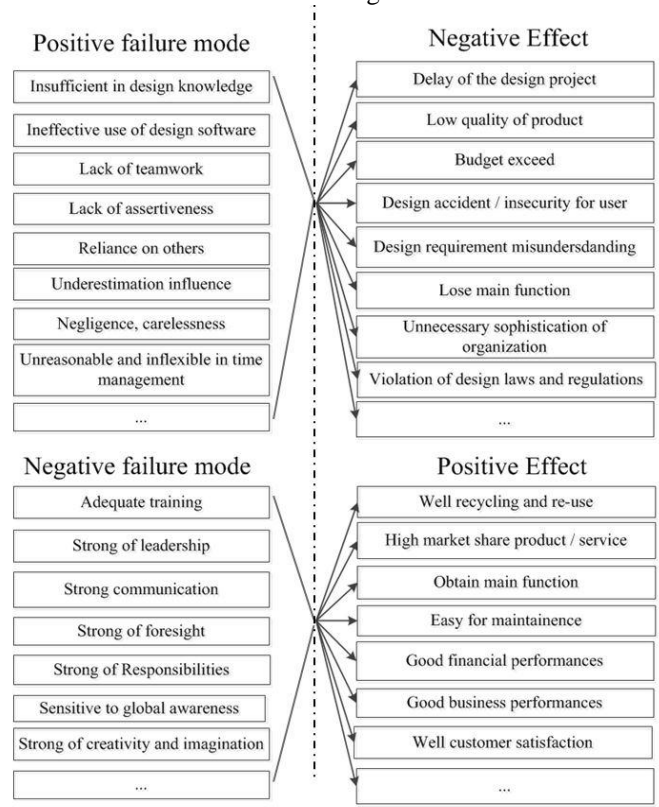

Fig. 6. Common failure and corresponding effect for designer

Depends on the effect of these failures in the actors, project manager can define the severity of these failures.

After brainstorming the failures and analysing the effects of these failures, project manager can use the evaluation criteria (Table I, Table II and Table III) for design process, to assign all the severity, detection and occurrence risk factor rank for failure existing actors (project manager can get the rank information through the interview with the candidate actors). To let FMEA methodology adapt our risk management method, we should, firstly, list all the failures for actor, and assign the rank to all these failures. After that, we can calculate total RPN (TRPN) for the failure existing actor.

A risk priority number (RPN) will be determined for each potential failure mode and effect, which by multiplying the ranking for the three factors (severity × occurrence × detection). FMEA hinged on Risk Priority Number for root causes of the potential failure modes to appraise the risk of the system and prioritise the actions that need to be taken [34]. Those failure modes with the highest RPNs should be attended to first, although special attention should be given when the severity ranking is high regardless of the RPN [31].

To exactly understand the process of TRPN calculation, we can make the total calculation process diagram like Fig. 7.

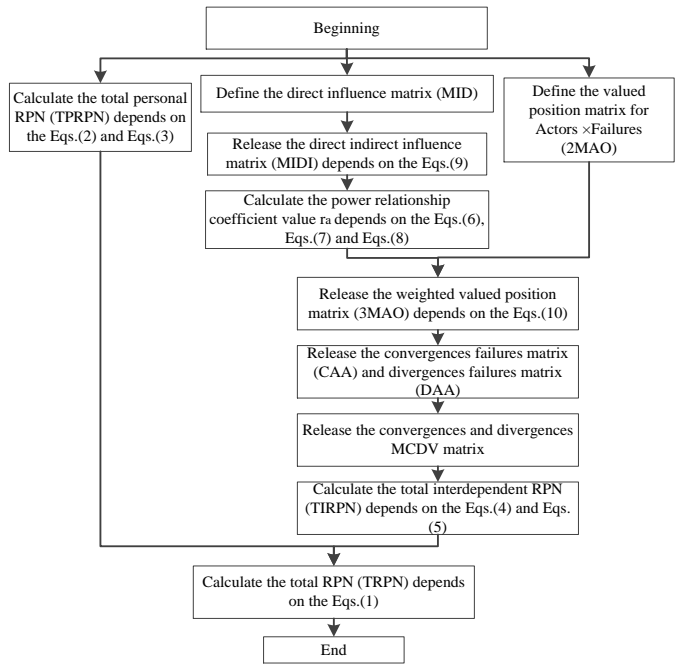

Fig. 7. TRPN calculation process

In the TPRN calculation process we can understand that when we calculate the total RPN for the failure existing actor, we should, firstly, calculate the personal RPN (Table IV) and total interdependent RPN respectively (Eqs. (4) and Eqs. (5)), and combine them together get final total RPN (Eqs. (1)).

In here, we can suppose that there are 10 actors, which with the name from Actor 1 to 10 and some of them have hidden failure (project manager can brainstorm the failure with the direction of process and organization, and release the corresponding effect like Fig. 6). Meanwhile, project manager can calculate PRPN for all the risk exiting actors (Table IV) depends on the Eqs. (2) and Eqs. (3).

TABLE IV. CALCULATION OF TOTAL PERSONAL RPN FOR FAILURE EXISTING ACTORS

| Failure existing actor | Failure mode | Effect | S | D | O | RPN | PRPN |
|---|---|---|---|---|---|---|---|
| Actor 1 | Lack of leadership | Delay of design project | 3 | 2 | 2 | 12 | $PRPN_1$ |
| | Lack of knowledge | Design accident/ insecurity for user | 1 | 3 | 1 | 3 | $PRPN_2$ |
| TPRPN | | | | | | 15 | |
| Actor 3 | Lack of Responsibilities | Delay of the design project | 2 | 5 | 3 | 30 | $PRPN_1$ |
| | Insensitive to global awareness | Low market share | 2 | 2 | 5 | 20 | $PRPN_2$ |
| | Poor communication | Low quality of product | 3 | 2 | 5 | 30 | $PRPN_3$ |
| TPRPN | | | | | | 80 | |
| Actor 6 | Lack of Responsibilities | Delay of the design project | 2 | 2 | 4 | 16 | $PRPN_1$ |
| TPRPN | | | | | | 16 | |
| Actor 7 | Poor communication | Low quality of product | 2 | 4 | 5 | 40 | $PRPN_1$ |
| TPRPN | | | | | | 40 | |
| Actor 8 | Lack of knowledge | Design accident/ insecurity for user | 1 | 2 | 3 | 6 | $PRPN_1$ |
| TPRPN | | | | | | 6 | |

Note. S: Severity. D: Detection. O: Occurrence. RPN = risk priority number. $PRPN_n$ = nth Personal Risk priority number. TPRPN = Total Personal RPN.

From the Table IV we can find that there are 5 failure existing actors (Actor1, Actor3, Actor6, Actor7 and Actor8) in here with the different failures in every actor. For the Actor1 in the Table IV, the values for PRPN1 (personal risk priority number for failure lack of leadership) and PRPN2 (personal risk priority number for the failure lack of knowledge) are 12 (3 × 2 × 2) and 3 (1 × 3 × 1). Therefore, the TPRPN (total personal priority number) is 15 (12+3). Meanwhile, in the Eqs. (2), the value p means the total number of existing failures for the actor. Depends on the Table IV, we can get TPRPN for all these five actors.

To adapt the FMEA method to the future point-to-point organization model, and considering about the direct, indirect, favorable and unfavorable effect, we should think about total interdependent RPN. The interdependent RPN is the risk priority number for the interdependent risk (Fig. 5). The total interdependent RPN (TIRPN) (Eqs. (4) and Eqs. (5)) is the risk priority number for all direct and indirect effected actors by main failure existing actor.

The following equations are the calculation of total RPN for the failure existing actor in the future point-to-point organization model.

$$TRPN = TPRPN + TIRPN \qquad (1)$$

$$TPRPN = \sum_{n=1}^{p}(PRPN_n) \qquad (2)$$

$$PRPN_n = S_n \times D_n \times O_n \qquad (3)$$

$$TIRPN = \sum_{n=1}^{M}(IRPN_n) \qquad (4)$$

$$IRPN_n = TPRPN \times MCDV_{a,n} \qquad (5)$$

Note. TRPN = Total RPN. TPRPN = Total Personal RPN. PRPNn = nth personal RPN. Sn = nth severity. Dn = nth Detection. On = nth Occurrence. TIRPN = Total interdependent RPN. IRPN n = the nth Interpersonal RPN. MCDVa,n = Convergence and Divergence value for failure in risk existing actor "a" with the nth effected actor.

The equation (4) means that we should calculate all the corresponding interdependent RPN (Eqs. (5)), and make sum to all these result to define the TIRPN. The M in the equation (4) is the total number of corresponding affect actors.

For the equation (5), we should introduce the new concept of MCDV (Matrix of Convergence and divergence value for failure). Thanks a lot for the MACTOR (Matrix of Alliances and Conflicts: Tactics, Objectives and Recommendations) [36] methodology, which is the meeting of the actors of the system according to their objectives, projects and means of action, allows revealing a number of strategic issues and to underline the key questions for the future on which they have convergence and divergence [33]. In here, the main objective of this method is to research the possible convergences and divergences of the different actors relative to the objectives of the project [37]. To adapt the MACTOR method to our failure interdependent actor relationship, we can replace the objectives in the method to the failures in the actor. Different from the convergence or divergence to objective between the two actors, in the MACTOR method, the convergence or divergence failure means the more failure infectious or less failure infectious to the other actors. The more convergence level of failure two actors have, the more failure two actors will infected with each other. In the 2MAO (valued positions matrix (Actors × Objectives)) in MACTOR method, positioning of actors in relation to objectives on a scale from -3 to 3, according to whether the level of opposition or agreement is high, medium or low [36]. Therefore, for the rank of the failure in the actor also can be a scale from -3 to 3, according to whether the failure level of positive (the actor is strong inclined to failure) or negative (the actor is strong inclined to defeat this failure) is high, medium or low. The example of the 10 actors valued position matrix for Actors × Failures can be seen as the Table V.

TABLE V. VALUED POSITION MATRIX FOR ACTORS × FAILURES (2 MAO)

| Actors | LL | LK | LR | PC | IGA |
|---|---|---|---|---|---|
| Actor 1 | 3 | 1 | 0 | 0 | 0 |
| Actor 2 | 0 | -1 | -2 | 0 | 0 |
| Actor 3 | 0 | 0 | 2 | 2 | 3 |
| Actor 4 | -2 | -1 | 0 | 0 | 0 |
| Actor 5 | -3 | 0 | 0 | 0 | 0 |
| Actor 6 | 0 | -1 | 2 | 0 | 0 |
| Actor 7 | 0 | 0 | 0 | 2 | 0 |
| Actor 8 | 0 | 1 | -2 | 0 | -1 |
| Actor 9 | 0 | 0 | -1 | 0 | 0 |
| Actor 10 | 0 | -2 | 0 | 0 | 0 |

Note. LL: Lack of leadership. LK: Lack of knowledge. LR: Lack of responsibility. PC = Poor communication. IGA = Insensitive to global awareness.

In the Table V, we can find that only Actor1, Actor3, Actor6, Actor7 and Actor8 have the positive value in one or several failures, which just match the failure existing actor in Table IV. Meanwhile, some actors have the negative value, such as the Actor 2 in the failure part of LK (Lack of knowledge), contains the value -1. It means that Actor2 can defend LN failure and have a positive effect to the risk, which means reduce the influence of the failure.

In the MACTOR method, before release the matrix of convergences and divergences (actors × actors), to reveal more real relationship among actors, it should be introduce the relationship of power between actors. In here, the more power of one actor contains means the more effect of failure influence to other actors. Thanks again to the MACTOR method to take into account the direct and indirect influences (Eqs. (9)) between two actors, which just match to our research of failure risk relationship between actors (Fig. 5), when considering about the power relationship among actors. In here, the indirect influence is exerted through the use of the influence with other intermediary actors [38]. The calculation process of power coefficient ($r_a$) for actor can be seen as the equation (6) to equation (9).

$$r_a = \frac{I_a - MIDI_{a,a}}{\sum_A I_a} \cdot \frac{I_a}{I_a + D_a} \qquad (6)$$

$$I_a = \sum_b (MIDI_{a,b}) - MIDI_{a,a} \qquad (7)$$

$$D_a = \sum_b (MIDI_{b,a}) - MIDI_{a,a} \qquad (8)$$

$$MIDI_{a,b} = MID_{a,b} + \sum_c \left(\min(MID_{a,c}, MID_{c,b})\right) \qquad (9)$$

Note. Ia = Net of direct and indirect influence of the actor "a". Da = Net of direct and indirect dependence of actor "a". MIDIa,b = direct and indirect influence from the actor "a" to actor "b".

In the equations above (Eqs. (6), (7), (8) and (9)), the MIDI means the matrix of direct and indirect influences. In addition, MID means the matrix of direct influence (Table VI). In the MACTOR method, the potential influence of one actor over another is recorded on a scale from 0 to 3 (none, weak, average, strong) [36]. In here, we can use MID matrix and Eqs(9) to calculate the MIDI matrix(Table VII).

TABLE VI. MATRIX OF DIRECT INFLUENCE MID

| To<br>From | A1 | A2 | A3 | A4 | A5 | A6 | A7 | A8 | A9 | A10 |
|---|---|---|---|---|---|---|---|---|---|---|
| Actor1 | 0 | 0 | 2 | 0 | 0 | 0 | 0 | 0 | 0 | 0 |
| Actor2 | 1 | 0 | 1 | 3 | 1 | 0 | 0 | 0 | 0 | 0 |
| Actor3 | 2 | 0 | 0 | 2 | 1 | 1 | 0 | 0 | 1 | 2 |
| Actor4 | 0 | 0 | 2 | 0 | 1 | 1 | 0 | 0 | 2 | 0 |
| Actor5 | 0 | 1 | 0 | 1 | 0 | 2 | 1 | 0 | 1 | 1 |
| Actor6 | 0 | 2 | 0 | 1 | 2 | 0 | 2 | 0 | 0 | 2 |
| Actor7 | 0 | 3 | 0 | 1 | 2 | 1 | 0 | 1 | 1 | 0 |
| Actor8 | 2 | 0 | 1 | 0 | 1 | 2 | 2 | 0 | 1 | 0 |
| Actor9 | 1 | 1 | 2 | 2 | 0 | 0 | 0 | 2 | 0 | 0 |
| Actor10 | 0 | 0 | 1 | 1 | 1 | 0 | 1 | 1 | 1 | 0 |

TABLE VII. MATRIX OF DIRECT AND INDIRECT INFLUENCE MIDI

| To<br>From | A1 | A2 | A3 | A4 | A5 | A6 | A7 | A8 | A9 | A10 | Ii |
|---|---|---|---|---|---|---|---|---|---|---|---|
| Actor1 | 2 | 0 | 2 | 2 | 1 | 1 | 0 | 0 | 1 | 2 | 9 |
| Actor2 | 2 | 1 | 4 | 5 | 3 | 3 | 1 | 0 | 4 | 2 | 24 |
| Actor3 | 3 | 3 | 6 | 6 | 4 | 3 | 3 | 2 | 5 | 4 | 33 |
| Actor4 | 3 | 3 | 4 | 6 | 3 | 3 | 2 | 2 | 4 | 4 | 28 |
| Actor5 | 2 | 5 | 4 | 6 | 6 | 4 | 4 | 3 | 4 | 3 | 35 |
| Actor6 | 1 | 5 | 3 | 6 | 7 | 4 | 4 | 2 | 4 | 3 | 35 |
| Actor7 | 3 | 6 | 4 | 7 | 6 | 5 | 3 | 2 | 4 | 2 | 39 |
| Actor8 | 4 | 6 | 4 | 5 | 6 | 5 | 5 | 2 | 4 | 4 | 43 |
| Actor9 | 6 | 1 | 7 | 5 | 4 | 4 | 2 | 2 | 4 | 2 | 33 |
| Actor10 | 3 | 3 | 4 | 5 | 5 | 5 | 3 | 3 | 6 | 2 | 37 |
| Di | 27 | 32 | 36 | 47 | 39 | 33 | 24 | 16 | 36 | 26 | 316 |

Depend on the MIDI in the Table VII and the equations from the Eqs. (6) to Eqs. (8), we can calculate power relationship coefficient $r_i$ for all the candidate actors. The Actor1, for instance, the value of the $I_{Actor\ 1}$ and $D_{Actor1}$ are $9((2+0+2+2+1+1+0+0+1+2) - 2)$ and $27 ((2+2+3+3+2+1+3+4+6+3) - 2)$. Then, depend on the Eqs. (6), the $r_{Actor1}$ is $0.00553$ $((((9-2)/(316)) \times (9/(9+27)))$. In here, we should consider the $r^*_{Actor1} = N \cdot \frac{r_i}{\sum ri}$ to be the final result of coefficient value for the reason of facilitate understanding and calculation. In here, the N means the number of the actor. Therefore, the result of the $r^*_{Actor1}$ is $0.12$ $(10 \times (0.00553/0.47151))$.

After we calculate the power coefficient value ($r_a$) for every actor, we multiply this value to the 2MAO for the failure (Table V) to get the 3MAO for failure (Eqs. (10)).

$$3MAO_{a,i} = 2MAO_{a,i} \cdot r_a \qquad (10)$$

Depends on the 3MAO, we can release 3CAA (valued matrix of convergences) and 3DAA (valued matrix of divergences). In here, for the reason of calculate the total failure influence from one actor to another actor, we can combine value of convergences and divergences, and define the final matrix of convergences and divergences of the actors for the failure (MCDV) (actors × actors). For the combination of the convergences and divergences, we should subtract value from the 3CAA to the 3DAA, then, divide by 9 (The absolute value of 3CAA and 3DAA level is range from 0 to 9). The value in the matrix of MCDV, concerning about the power effect and common failure level from one actor to the other actor, which will be the interdependent effect level from one actor to another actor. The result matrix for MCDV can be seen as the Table VIII.

TABLE VIII. MCDV FOR 10 CANDIDATE ACTORS

| To<br>From | A1 | A2 | A3 | A4 | A5 | A6 | A7 | A8 | A9 | A10 |
|---|---|---|---|---|---|---|---|---|---|---|
| Actor1 | 0.05 | -0.04 | 0 | -0.12 | -0.17 | -0.06 | 0 | 0.12 | 0 | -0.16 |
| Actor2 | -0.04 | 0.22 | -0.16 | 0.06 | 0 | -0.08 | 0 | 0.15 | 0.12 | 0.19 |
| Actor3 | 0 | -0.16 | 0.67 | 0 | 0 | 0.21 | 0.27 | -0.58 | -0.14 | 0 |
| Actor4 | -0.12 | 0.067 | 0 | 0.19 | 0.21 | 0.09 | 0 | -0.14 | 0 | 0.19 |
| Actor5 | -0.18 | 0 | 0 | 0.21 | 0.31 | 0 | 0 | 0 | 0 | 0 |
| Actor6 | -0.07 | -0.09 | 0.21 | 0.09 | 0 | 0.36 | 0 | -0.51 | -0.17 | 0.21 |
| Actor7 | 0 | 0 | 0.27 | 0 | 0 | 0 | 0.33 | 0 | 0 | 0 |
| Actor8 | 0.12 | 0.16 | -0.58 | -0.14 | 0 | -0.51 | 0 | 0.89 | 0.28 | -0.27 |
| Actor9 | 0 | 0.12 | -0.14 | 0 | 0 | -0.17 | 0 | 0.28 | 0.1 | 0 |
| Actor10 | -0.16 | 0.19 | 0 | 0.19 | 0 | 0.21 | 0 | -0.27 | 0 | 0.31 |

Depends on the Table VIII, we can understand the failure risk effect relationships between Actor 1 (failure existing actor) and other actors (Fig. 8).

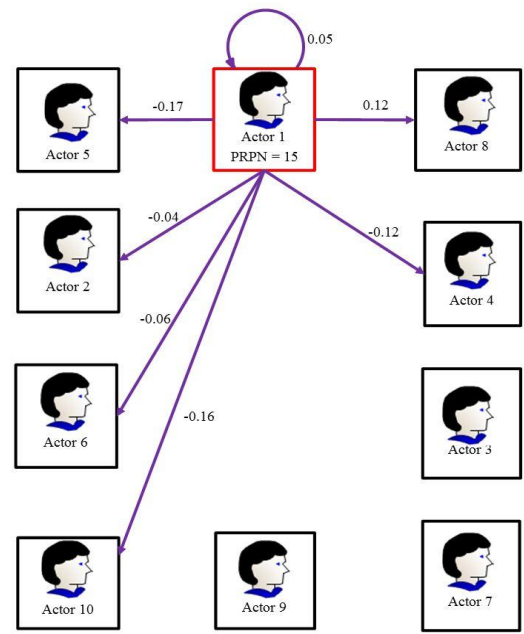

Fig. 8. Actor 1 hidden risk relationships

In the Fig. 8, we can understand that the actor 1 will effect Actor 5, Actor 2, Actor 6, Actor 10 and Actor 8 with the different influence value. In the Fig. 8, the arrow means the combination of direct and indirect effect in the Fig. 5. In here, in the Eqs. (5), the reason why multiply MCDVa,n with the PRPN is that when considering about the interdependent RPN we should use the weight of interdependent effect and Personal RPN to define the final interdependent RPN. After that, we can combine all these interdependent effects to release the total interdependent RPN (Eqs. (4)). In the Fig. 8, we can calculate the TIRPN for Actor 1 is -5.7 (depends on the Eqs. (4) and Eqs. (5)). Hereafter, depends on the equation (1) to calculate Total RPN for all the failure existing actors, project manager can prioritize the failure actors for action. The higher value of Total RPN the actor contains, the more priority level for the actor has. Therefore, TRPN for the Actor1 in the Fig. 8 is 9.3(15 + (-5.7)). The result TRPN for all these failure actors can be seen as the Table IX.

TABLE IX. TOTAL RPN RESULT FOR ALL THE HIDDEN RISK EXISTING ACTORS

| Failure existing actor | TRPN |
|---|---|
| Actor 1 | 10 |
| Actor 3 | 102 |
| Actor 6 | 16 |
| Actor 7 | 64 |
| Actor 8 | 6 |

In the Table IX, we can find that the priority order for action actor is Actor 3 (102), Actor 7(64), Actor 6(16), Actor 1 (10) and Actor 8(6). Project manager can depends on this order to take action (training, course, education and etc) to reduce the risk for the failure actors. After that, if the risk is still not tolerable, the failure actor will be eliminated, otherwise, complete the organization and do the final allocation.

## IV. CONCLUSIONS

In this paper, we firstly introduce the existing problems for the future organization structure, including the complexity of analysing the risk for the future organization. Then, main objective is illustrated to confront with the problems mentioned above. After that, we propose a methodology to approach the existing problems, especially use the FMEA method to evaluate the personal risk and use the MACTOR method to evaluate the interdependent risk for the future organization structure.

From here, we can integrate all these information to the P-P-O model, and let it adapt much more to the factory of the future.